\documentclass[slac_one]{revtex4}
\usepackage{graphicx}
\usepackage{fancyhdr}
\newcommand{\jpsi}{{\ensuremath{J/\psi}}}
\newcommand{\ups}{{\ensuremath{\Upsilon(1S)}}}
\newcommand{\upss}{{\ensuremath{\Upsilon(2S)}}}
\newcommand{\upsss}{{\ensuremath{\Upsilon(3S)}}}

\newcommand{\ptjpsi}{{\ensuremath{p_{T}^{J/\psi}}}}
\newcommand{\ptups}{{\ensuremath{p_{T}^{\Upsilon}}}}

\begin{document}

\title{Perspectives on Heavy Flavour Production and Spectroscopy Studies at the LHC} 

%

\author{A.~C.~Kraan\footnote{Financed by the European Union as Marie Curie fellow under contract number EIF PHY-040156.}, on behalf of the ATLAS and CMS Collaborations}
\affiliation{Istituto Nazionale di Fisica Nucleare di Pisa, Largo Pontecorvo 3, 56126, Pisa, Italy}
\begin{abstract}
In this note we discuss prospects of analyses at the ATLAS and CMS experiments that are related to the production of heavy flavour quarks. Trigger strategies are summarized and a selection of studies is presented. These include quarkonium production studies such as the differential transverse momentum cross section measurement and polarization studies, as well as reconstruction of excited quarkonium states. Furthermore inclusive $b$-quark production is discussed. 
\end{abstract}

\maketitle

\thispagestyle{fancy}


\section{INTRODUCTION} 
The LHC will provide new opportunities to explore the physics of heavy flavour quarks. 
This note focuses on analyses in CMS and ATLAS that are dedicated to the understanding of heavy flavour production. Studies on heavy flavour decays are summarized in~\cite{maria}. ATLAS and CMS 
are expected to contribute to the study of heavy quark physics thanks to their excellent capabilities to detect muons up to large pseudorapidity ($|\eta|<2.5$) and transverse momentum. Since charm and beauty quarks will be produced in abundance, many analyses dedicated to heavy flavour production are viable already with small integrated luminosity. In the results presented below, collisions are assumed to occur at 14 TeV centre-of-mass energy.

The outline of this note is as follows. In Section 2 trigger strategies for heavy flavour physics analyses will be outlined. Section 3 is devoted to quarkonium studies, including the \jpsi~cross section measurement, quarkonium polarization and reconstruction of $\chi_c$-states.  Similar analyses are being done in both experiments, but here we discuss only one analysis of each type. 
In Section 4 we discuss inclusive $b$-production. Section 5 contains conclusions.
\section{HEAVY FLAVOUR TRIGGERS}\label{trig}
The LHC bunch crossing rate at ATLAS and CMS at nominal luminosity will be 40 MHz. Via a Level 1 (L1) and a high level trigger (HLT) system the rate will be reduced to $\sim 200$ Hz before the events are stored~\cite{cmstdr,atlastdr}. 

The relevant triggers for heavy flavour analyses in CMS are:

\vspace*{-0.12cm}
\begin{itemize}
\item Dimuon trigger. There are triggers with invariant mass cuts for selecting certain resonances (\jpsi's, $\psi(2S)$, \ups, \upss, \upsss) and triggers without invariant mass cuts (e.g. for $B_s\rightarrow \mu^+\mu^-$). The L1 and HLT trigger muon $p_T$ threshold for both muons is 3 GeV/c (the $2\mu 3$ menu).
In addition there is a HLT menu based on selecting events with a displaced dimuon vertex, for collecting events containing \jpsi's produced in B-hadron decays.
\vspace*{-0.12cm}
\item Single muon trigger, where both at L1 and at HLT one muon is required to exceed a given $p_T$. Different trigger $p_T$ thresholds are anticipated and the lowest thresholds will be prescaled triggers.
\end{itemize}

In ATLAS, two trigger scenarios are considered. 
\vspace*{-0.12cm}
\begin{itemize}
\item Dimuon trigger. Two identified muons are required with a minimum $p_T$ threshold for the first and second muon of 6 and 4 GeV/c respectively ($\mu 6 \mu 4$).
At startup, the threshold of the $p_T$ of the first muon may be lowered to 4 GeV ($\mu 4\mu 4$). These menus are aimed at selecting low-mass resonances and non-resonance dimuon events.  
\vspace*{-0.12cm}
\item Single muon trigger. Here only one identified muon is required with a threshold of $p_T>10$ GeV/c ($1\mu 10$). For quarkonium analyses, a second muon is then searched for offline among the other tracks reconstructed in the Inner Tracker. The single muon trigger can also be used for heavy flavour analyses with hadronic final states, or final states with electrons, jets or photons.
\end{itemize}
\section{QUARKONIUM PRODUCTION STUDIES}
Inclusive quarkonium production at hadron colliders has been studied extensively in the theory of NRQCD~\cite{here}. One of the innovations of NRQCD is the so-called colour-octet mechanism (COM), which successfully explains the inclusive quarkonium cross section data at the Tevatron. However, recent polarization measurements, which revealed unpolarized \jpsi's~\cite{cdf}, are in clear contrast with COM predictions.
%
The colour singlet model (CSM) has recently been revived~\cite{maltoni}, but whether it can account for both cross section and polarization is still uncertain. 
In view of this puzzling situation and given the fact that quarkonia can be probed up to very large transverse momenta and with high statistics, it is very important to study quarkonium production at ATLAS and CMS. Quarkonia are additionally relevant because they are crucial for detector alignment and calibration. 
Below we discuss feasibility studies on quarkonium production at CMS and ATLAS.
\subsection{J/\boldmath{$\psi$} CROSS SECTION MEASUREMENT}
Three processes dominate \jpsi~hadro-production: prompt \jpsi's produced directly, prompt \jpsi's produced indirectly (via decay of heavier charmonium states such as $\chi_c$), and non-prompt \jpsi's from the decay of a B-hadron. CMS proposes a measurement of the \jpsi$\rightarrow\mu^+\mu^-$ differential cross section as a function of the transverse momentum of the \jpsi, \ptjpsi, with 3 pb$^{-1}$ of data~\cite{cmspas}. 

\begin{figure*}[h!]
\centering
\includegraphics[width=58mm]{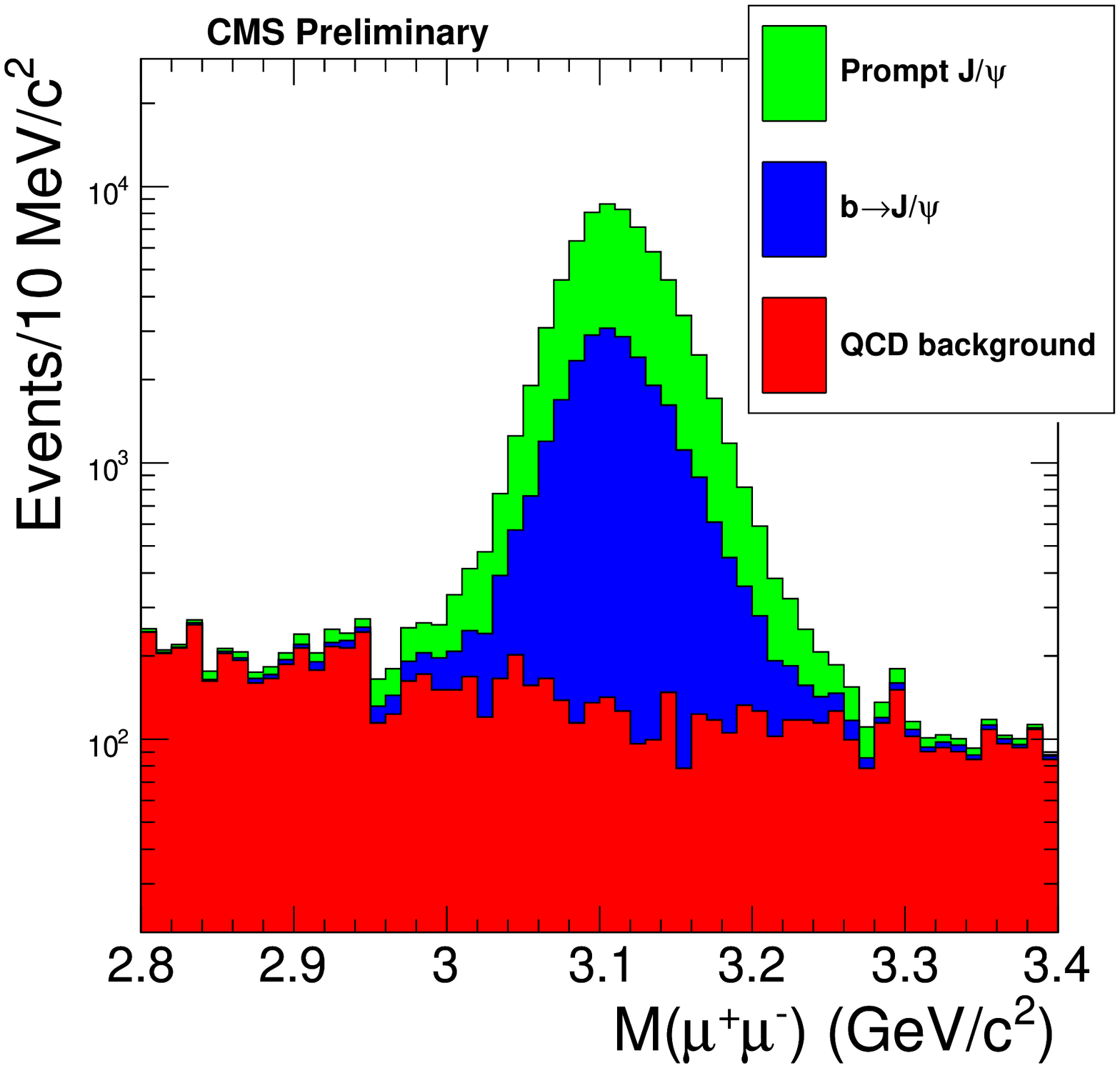}\includegraphics[width=58mm]{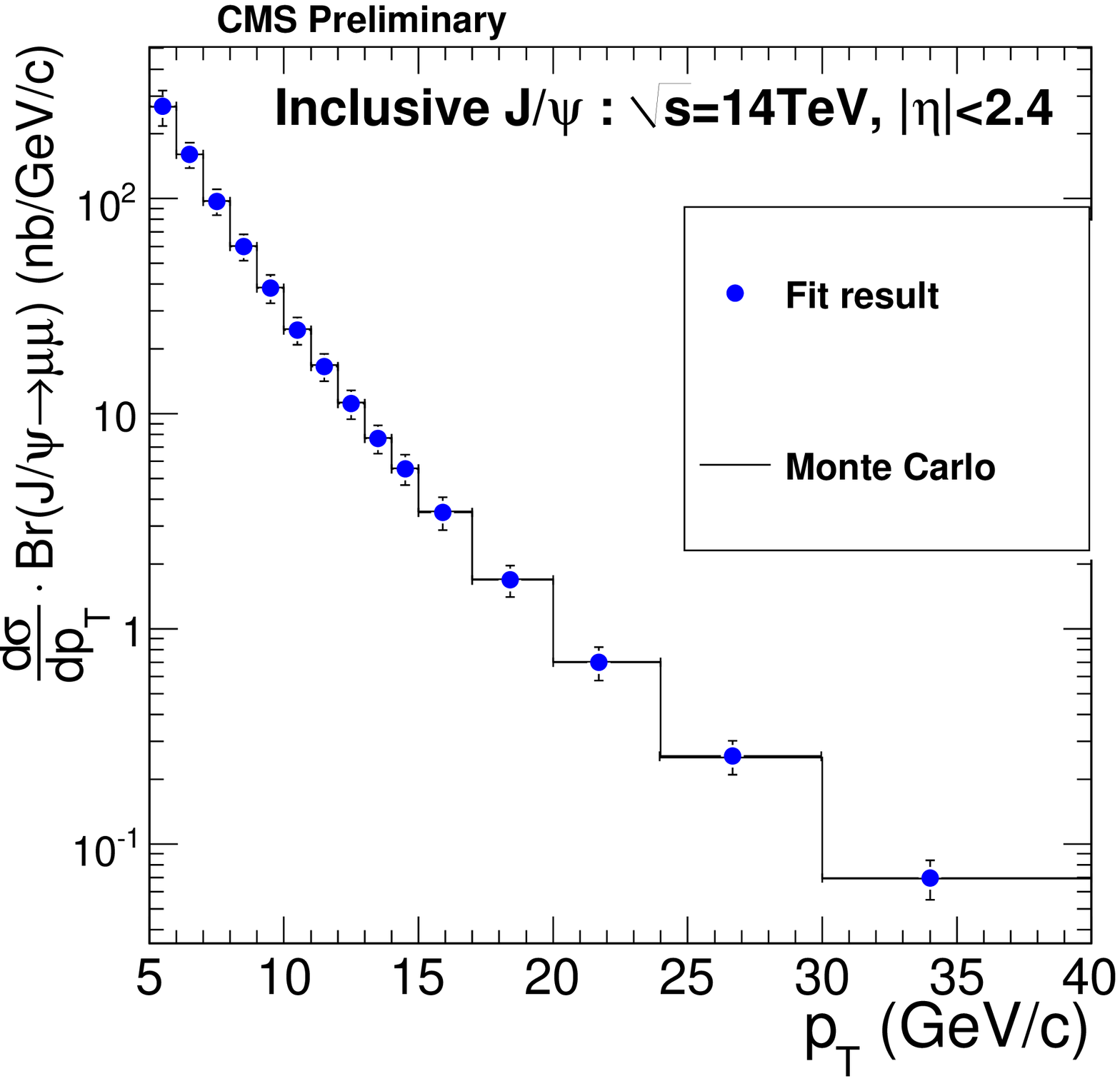}\includegraphics[width=58mm]{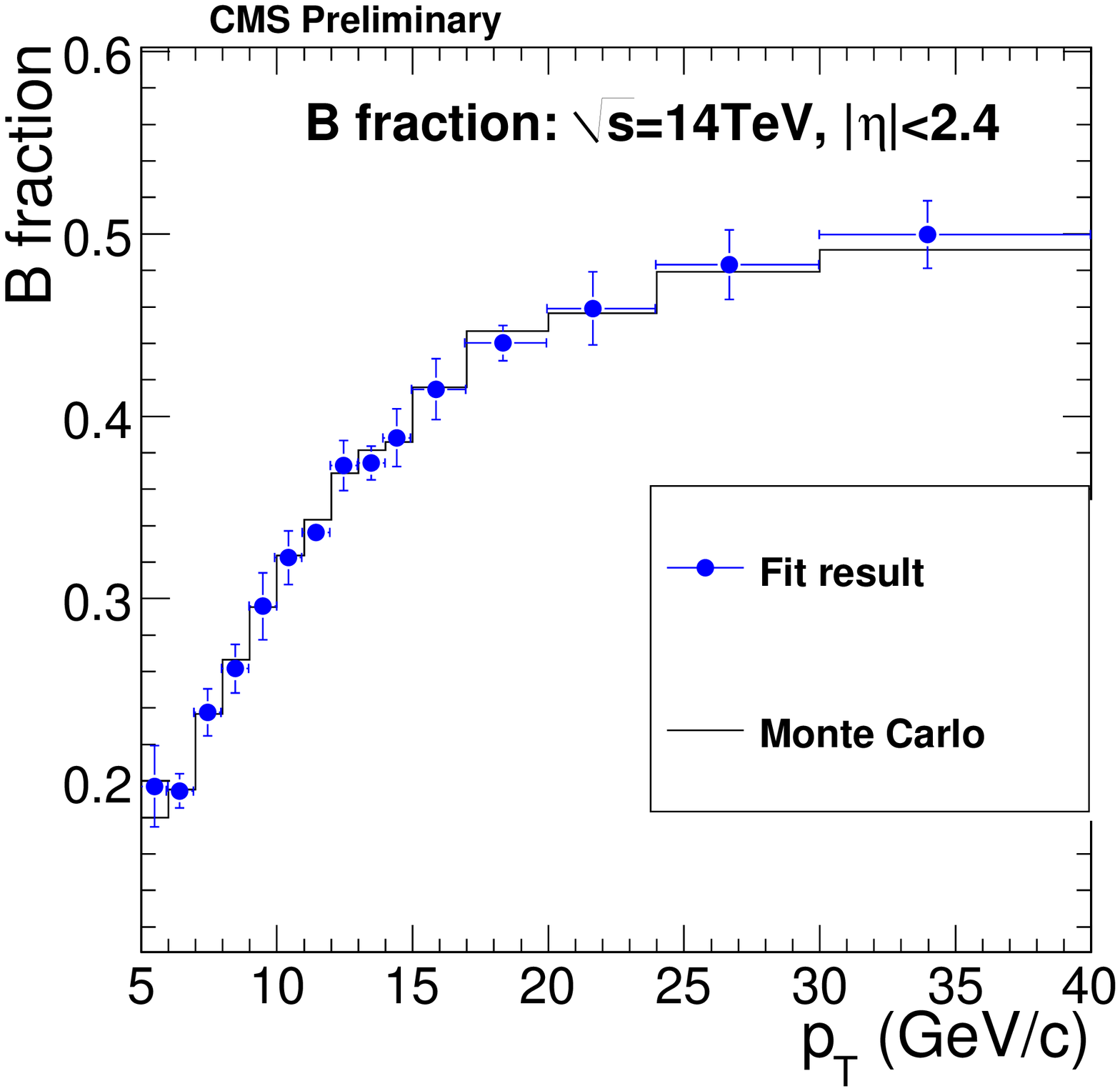}
\vspace*{-0.2cm}
\caption{Left: invariant mass distribution for prompt \jpsi's, non-prompt \jpsi's and background. Middle: inclusive \jpsi~differential cross section. Right: fraction of \jpsi's from B-hadron decay. All plots are based on expectations with 3 pb$^{-1}$ of CMS data.} \label{ref1}
\end{figure*}

Prompt and non-prompt \jpsi~events will be selected with the $2\mu 3$ trigger, followed by a set of offline selection cuts including invariant mass and dimuon vertex requirements. The main background comes from any other source of muons that, when paired, could accidentally have an invariant mass close to that of the \jpsi, i.e. muons from heavy flavour meson decay and $K/\pi$ decays-in-flight.
It must be noted that CMS cannot reconstruct \jpsi's with a transverse momentum below about 5 GeV/c. Fig.~\ref{ref1} (left) displays the invariant mass for prompt \jpsi's, non-prompt \jpsi's, and background. The yield per 1 pb$^{-1}$ of data is expected to be 25000 \jpsi's and the mass resolution averaged over all $\eta$ values is roughly 30 MeV/$c^2$.

The inclusive \jpsi~differential cross section will be measured by fitting the mass spectrum with a \jpsi~signal and a background hypothesis. 
The fraction $f_B$ of $J/\psi$'s from B-hadron decays in this study is determined by using decay length distributions. In each bin of \ptjpsi~ an unbinned maximum-likelihood fit to the data is performed. Dominant systematic errors are the luminosity and the polarization uncertainties. The inclusive \jpsi~differential cross section and the fraction of \jpsi's from B-decay with 3 pb$^{-1}$ of data are displayed in Fig.~\ref{ref1} (center and right). 

\vspace*{-0.1cm}
\subsection{QUARKONIUM POLARIZATION MEASUREMENT}
ATLAS has outlined a proposal for measuring the polarization of \jpsi~and \ups~states~\cite{atlasnote}.  The \jpsi~polarization (naturally the same holds for all quarkonia) can be inferred from the angular distribution of the muons from \jpsi~decay. Looking in the \jpsi~rest frame, $\theta$ is defined as the angle between the $\mu^+$ and the direction of the motion of the \jpsi~in the laboratory frame. The normalized angular distribution is $I(\cos\theta)=\frac{3}{2(\alpha+3)}(1+\alpha\cos^2\theta)$. Unpolarized, transversely and longitudinally polarized \jpsi's have $\alpha=0$, $\alpha=1$ and $\alpha=-1$, respectively. 

\begin{figure*}[h!]
\centering
\vspace{-4mm}
\includegraphics[width=69mm]{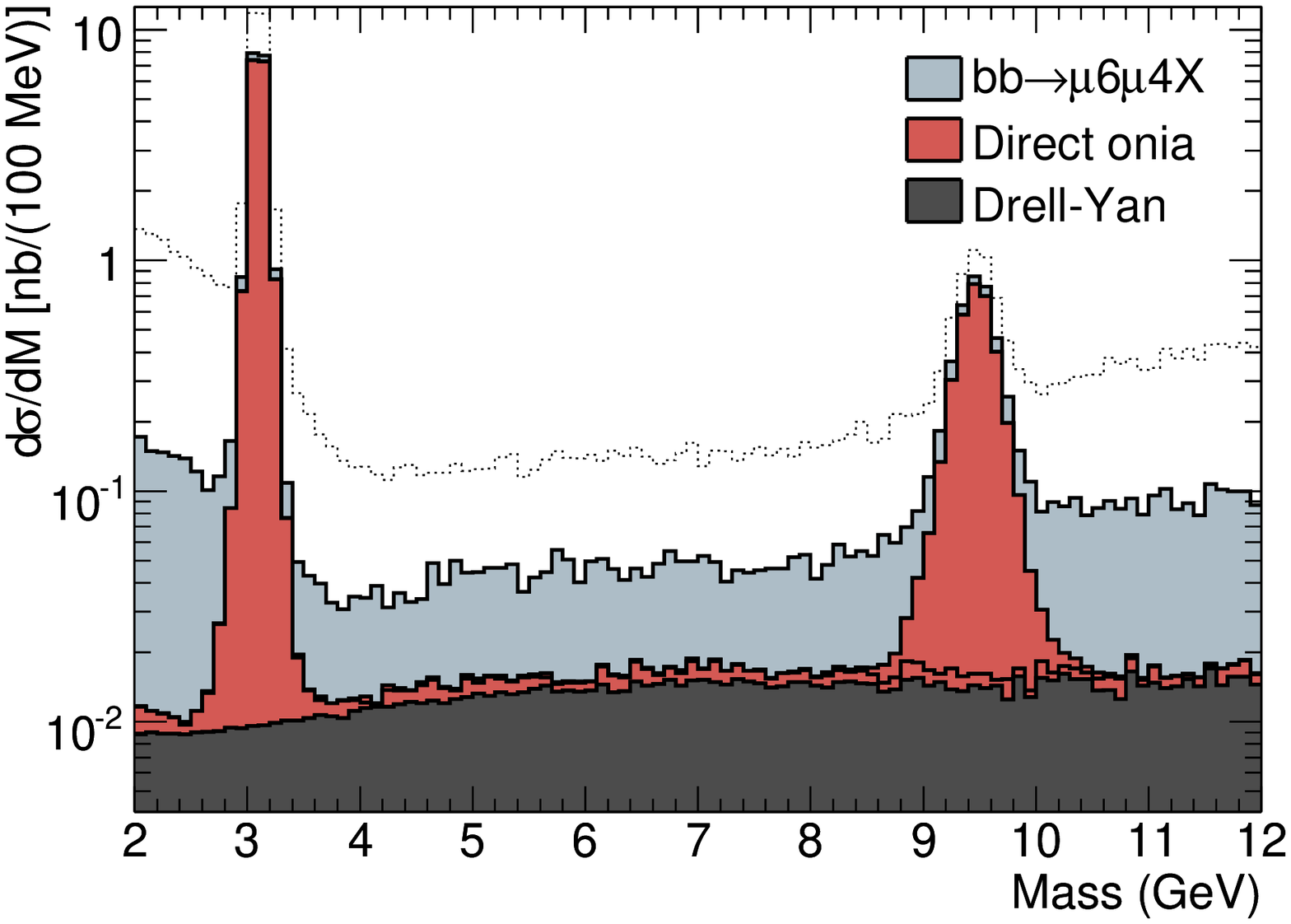}
\includegraphics[width=69mm]{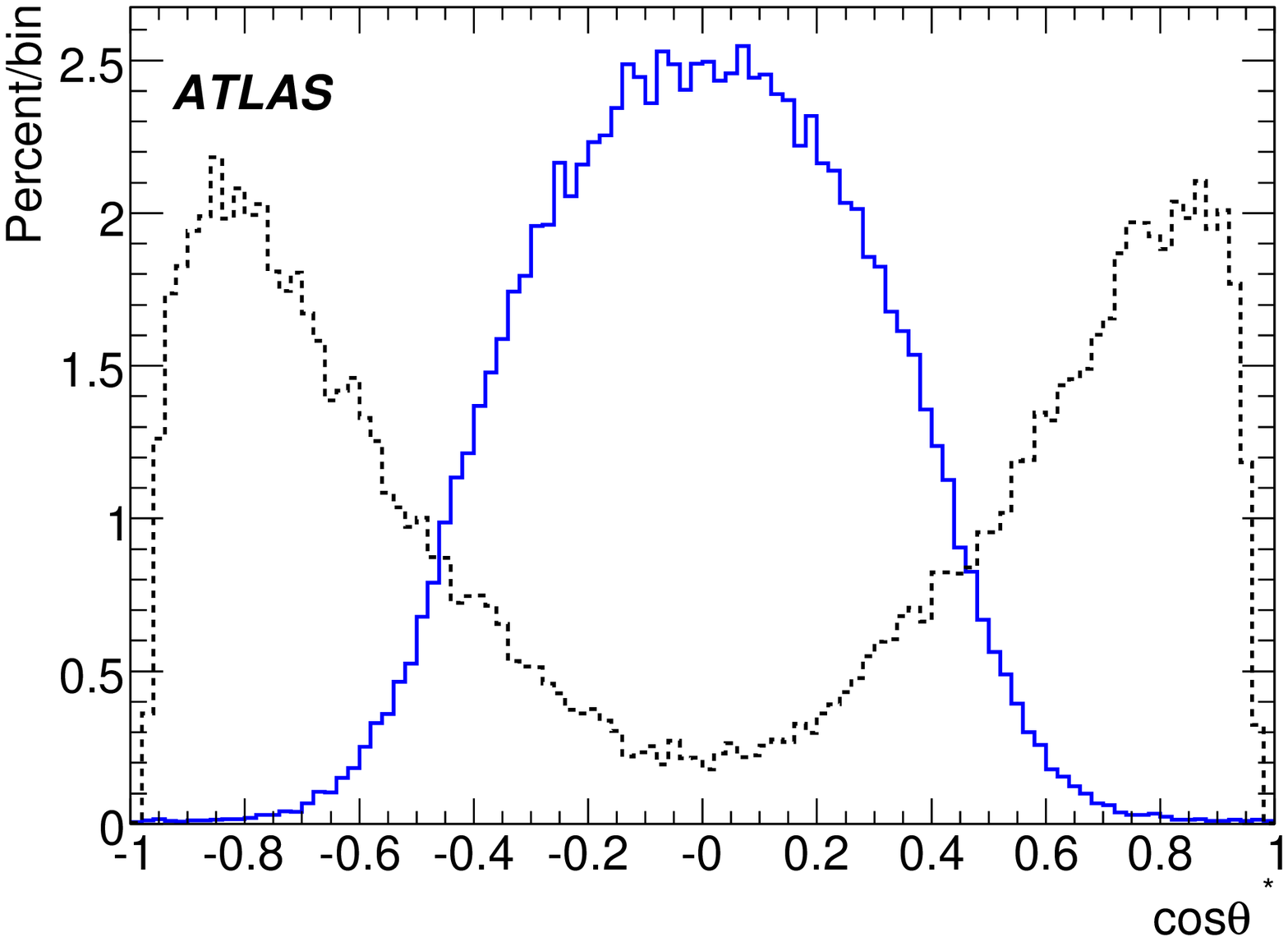}
\vspace*{-0.5cm}
\caption{Left: dimuon mass distribution expected with 10 pb$^{-1}$ of ATLAS data in the \jpsi~and \ups~mass region. The dashed line corresponds to omitting a vertex cut. Right: Reconstructed $\cos\theta$ distribution for the $\mu 6\mu 4$ dimuon trigger (solid line) and a $1\mu 10$ single muon trigger (dashed line). The true angular distributions are in both cases flat.} \label{ref2}
\end{figure*}

Quarkonium events will be selected with the $\mu 6\mu 4$ trigger (at startup possibly the $\mu 4\mu 4$ trigger). Offline vertex cuts and invariant mass cuts will be made. Due to detector effects no \jpsi's can be reconstructed below \ptjpsi$\simeq 8$ GeV/c. For \ups's the corresponding threshold is about 2 GeV/c. With the $\mu 4\mu 4$ startup menu \ups's could be reconstructed down to zero \ptups. The invariant mass distribution for \jpsi's and \ups's is given in Fig.~\ref{ref2}(left). The mass resolutions for \jpsi's and \ups's are expected to be about 54 and 170 MeV/$c^2$, respectively. The yield for 1 pb$^{-1}$ of data is 15000 \jpsi's and 4000 \ups's.

The reconstructed polarization angle for the $\mu 6\mu 4$ is displayed in Fig.~\ref{ref2}(right). The acceptance is maximal in the region where $|\cos\theta|\simeq 0$, corresponding to the case where the two muons have roughly equal transverse momentum. In the region $|\cos\theta|\simeq 1$ instead, the acceptance is zero. The reason is that here one muon is boosted and the other anti-boosted by the quarkonium, the latter not passing the dimuon trigger threshold. Hence ATLAS anticipates to use additionally the $1\mu 10$ trigger, and to reconstruct the \jpsi~by searching offline for a second muon track (see Sec.~\ref{trig}). This allows the slow muon to be detected only in the tracker, without having to pass any muon trigger thresholds, so sensitivity at $|\cos\theta|\simeq 1$ is recovered, see Fig.~\ref{ref2}~(right). Using the $1\mu 10$ trigger implies more background. While for \jpsi's the background is manageable, for \ups's the background is too high and the single muon trigger cannot be used. The sensitivity to measure the polarization in the $p_T$ range 12-20 GeV/c is 0.02-0.06 for \jpsi's and about 0.2 for \ups's with 10 pb$^{-1}$ of data. 

\vspace*{-0.1cm}
\subsection{\boldmath{$\chi_c$} RECONSTRUCTION}
ATLAS has studied the reconstruction of the excited quarkonium states $\chi_c$ and $\chi_b$~\cite{atlasnote}. 
Based on the fact that the angle $\alpha$ between the photon and the \jpsi~is very tiny, a photon is searched for in a very narrow cone around the \jpsi ($\cos\alpha>0.97$) and the $\mu\mu\gamma$ invariant mass is calculated. If the mass difference between the $\mu\mu$ and the $\mu\mu\gamma$ system is between 200 and 800 MeV/$c^2$ the system is considered to be a $\chi$ candidate. By fixing the masses of the $\chi_{c_0}$, $\chi_{c_1}$ and $\chi_{c_2}$ to their expected value, their relative contributions can be measured. 
\section{INCLUSIVE B-QUARK PRODUCTION}
\begin{figure*}[h!]
\centering
\vspace*{-0.5cm}
\includegraphics[width=7cm,height=5cm]{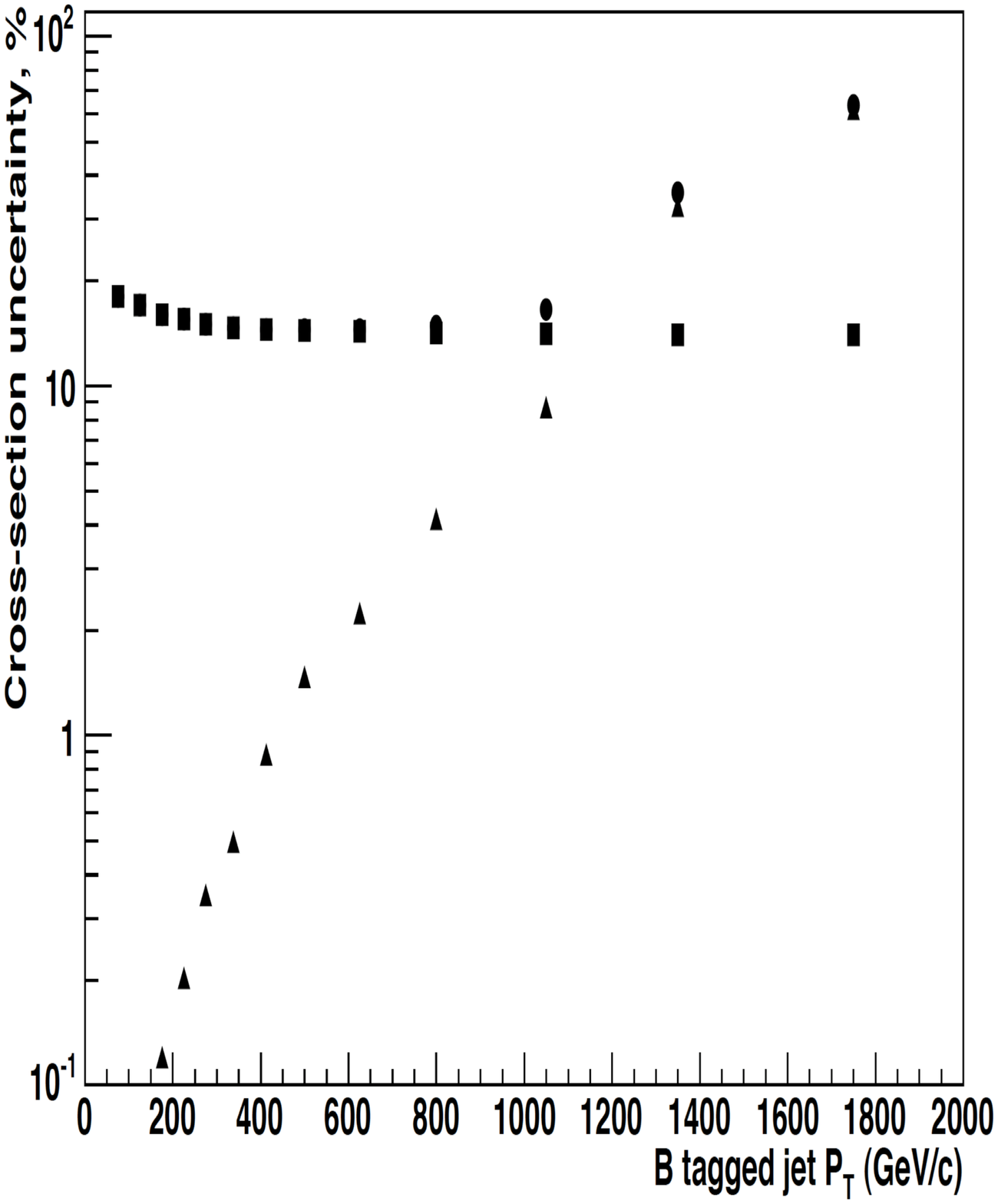}
\vspace*{-0.8cm}
\caption{
The expected statistical (triangles), systematic (squares) and total (dots) uncertainties on the cross section as function of the $p_T$ of the B-tagged jet expected at CMS. } \label{ref3}
\end{figure*}

B-production will be an abundant source of hadron production at the LHC.  
To disentangle the different contributions to production ($gg$ fusion and $q\bar{q}$ annihilation, flavour excitation, gluon splitting), measuring the $B$-hadron $p_T$ spectrum up to large momentum and with large statistics is important at the LHC. In CMS~\cite{bbbar}, B-hadron events will be selected with a L1 single muon trigger and a 'muon+$b$-jet' HLT. Offline b-jet tagging is applied. Each reconstructed muon is associated with the most energetic $b$-tagged jet, and must be closest to this jet than to any other jet. The transverse momentum of the muon is calculated with respect to the b-jet axis, which discriminates effectively between signal and background, and thereby allows to determine the signal fraction from a fit to the data.
In 10 fb$^{-1}$ of data about 16 million $b$-quark events are expected to be collected. The expected cross section uncertainty as a function of the $p_T$ of the $b$-tagged jet is given in Fig.~\ref{ref3}. 

\vspace*{-0.2cm}
\section{CONCLUSION}
We have summarized a selection of analyses expected to be performed at CMS and ATLAS in the context of heavy flavour production. More analyses are under study and first results based on real data are expected in 2009. 

\begin{acknowledgments}
Thanks to Zongchang Yang and Vato Kartvelishvili for discussions when preparing the talk, and to Fabrizio Palla and Carlos Lourenco for comments on this manuscript. 
\end{acknowledgments}

\end{document}